\documentclass[12pt,a4paper]{article}
\usepackage[utf8]{inputenc}
\usepackage[left=2.75cm,right=2.75cm,top=3cm,bottom=3cm]{geometry}
\usepackage{amsmath}
\usepackage{amsfonts}
\usepackage{amssymb}
\usepackage{amsthm}
\usepackage{bbm}
\usepackage{bm}
\usepackage{mathtools}
\usepackage{graphicx}
\usepackage{caption}
\usepackage[flushleft]{threeparttable}
\usepackage[dvipsnames]{xcolor}
\usepackage{booktabs}
\usepackage{rotating}
\usepackage{eurosym}
	\DeclareUnicodeCharacter{20AC}{\euro}
\usepackage{multirow}
\usepackage{longtable}
\usepackage{setspace}
\usepackage{relsize}
\usepackage{graphicx,subcaption}
\usepackage{pdflscape} 
\usepackage{xurl}
\usepackage{ragged2e}
\usepackage{float}
\usepackage[colorlinks=true]{hyperref}
	\hypersetup{citecolor=MidnightBlue, anchorcolor=Black, urlcolor=MidnightBlue, linkcolor=MidnightBlue}
\usepackage[nottoc]{tocbibind}
\usepackage{comment}
\usepackage[super]{nth}
\usepackage{tikz}
\usetikzlibrary{fit, calc, graphs, graphs.standard, decorations.pathreplacing, quotes,arrows.meta, positioning}
\usepackage{pgfplots} 				
	\usepgflibrary{shapes.geometric}
	\usepgfplotslibrary{polar}		
	\pgfplotsset{compat=1.18}	
	\tikzstyle{block2} = [rectangle, draw, fill=BrickRed!50, text width=8em, text centered, rounded corners=5, solid, minimum height=2em]
\usepackage[backend=biber, style=apa, bibstyle=authoryear, dashed = false, maxcitenames=2, maxbibnames=6, giveninits=true, sorting=nyt, date=year, eprint = false, doi=false, isbn=false]{biblatex} 
\newtheorem{proposition}{Proposition}

\newcommand\fnote[1]{\captionsetup{font=small}\caption*{#1}}

\title{La Révolution Dévore ses Enfants:\\
Pricing Implications of Transformative Agreements\footnote{I am grateful for valuable feedback from Iwan Bos, Oliver Budzinski, Justus Haucap, Ulrich Herb, Leon Knoke, Anastasiia Parakhoniak, Maikel Pellens, Michael Rose, Chris Snyder, and Christian Wey, as well as from participants of the \nth{11} OLIGO Workshop 2023 and the \nth{58} Hohenheim Colloquium 2023. All errors and opinions are my own. Funding by the German Research Foundation (DFG) is gratefully acknowledged (Funding No.~\#235577387/GRK 1974).}
}

\author{W.~Benedikt Schmal\thanks{Ilmenau University of Technology, Economic Theory Group, Ehrenbergstr.~29, 98693 Ilmenau, Germany \& KU Leuven University, Department for Management, Strategy, and Innovation (MSI), Naamsestraat 69, 3000 Leuven, Belgium. Email: \href{mailto:wolfgang-benedikt.schmal@tu-ilmenau.de}{wolfgang-benedikt.schmal@tu-ilmenau.de}.\\ ORCiD: \href{https://orcid.org/0000-0003-2400-2468}{0000-0003-2400-2468}}\\
\normalsize TU Ilmenau \& MSI, KU Leuven
}

\date{\emph{\today}}
\begin{document}
\begin{onehalfspace}
\maketitle
\thispagestyle{empty}

\vspace{-5mm}




\begin{center}
\emph{Fully revised version of \href{https://arxiv.org/abs/2403.03597v1}{``The 'Must Stock' Challenge in Academic Publishing:\\ Pricing Implications of Transformative Agreements''}}
\end{center}

\begin{abstract}
With the widespread dissemination of the internet, academia envisioned free availability and rapid dissemination of new knowledge. However, most researchers continued publishing in established journals instead of switching to fully open-access alternatives. 
That preserved the market power of the large commercial publishing houses owning thousands of journals behind paywalls. To turn these portfolios into open access, research institutions around the globe have been negotiating `transformative agreements:' Papers are published fully open access, and universities pay only for the publication but not for subscriptions any longer. In this paper, I demonstrate that publishers controlling a large stock of paywalled publications can use them as leverage to ensure high revenues even with decreasing publication numbers. By that, this industrial policy may harm competitors that only publish under open access. This could harm competition and perpetuate the position of the incumbent players.
\end{abstract}

\vspace{5mm}

\end{onehalfspace}
\newpage

\begin{doublespace}
\section{Introduction}
\label{sec.intro}

\emph{``The revolution devours its children''} diagnosed Jacques Mallet du Pan in 1793, reflecting on the evolution of the French Revolution since 1789. After the monarchy had been overthrown in France, things did not change in the way many people envisioned it. In academic publishing, the internet made the ideal of free and unlimited sharing of new discoveries, insights, and methods graspable. In parallel, libraries faced increasing pressure on their budgets as subscription costs grew starkly \parencite{Bergstrom.2001}. Publishers exploited the market power of their flagship journals \parencite{Dewatripont.2007} by selling them only in massive bundles \parencite{Bergstrom.2014},.

The situation was fertile ground for establishing `gold' open access: Journals that do not charge any subscriptions but a publication fee for accepted articles. In turn, every article is published freely accessible online. While few journals performed well, the overall market remained inert -- shifting towards new outlets is a classic collective action problem \parencite[][]{Bergstrom.2001}. However, the idea of ungated access to journal publications sparked interest. Publishers with huge subscription-based journal portfolios began to sell open access to individual articles to the submitting authors. It allows them to publish in their familiar, highly ranked incumbent journals but still contribute to open science \parencite{Schmal.2023a} and increase citations in disciplines without a preprint culture \parencite{MuellerLanger.2018}. The downside is that research institutions pay twice: Once for the subscriptions that continued to be charged and for open access purchased by their affiliated researchers \parencite{Laakso.2016}.

Accepting that switching to a fully open access publishing landscape without the journal portfolios of the legacy publishers would be unrealistic, libraries began to negotiate large-scale `transformative agreements' (TAs). The present paper is the first to model publisher behavior in such a contract theoretically. It reveals that legacy publishers could `insure' themselves against declining submissions. By that, TAs may absorb budgets that cannot be used for spending on gold open access: The open science transformation may economically harm its (gold open access) children. 

Transformative agreements are widespread. In August 2024, more than 1,000 TAs were registered in the public ESAC database \parencite{Rothfritz.2024}. In 2020, some 100,000 articles were published under TA. It grew exponentially to some 600,000 in 2022 and nearly 900,000 publications in 2023.\footnote{See the data from the ESAC Initiative, \url{https://esac-initiative.org/market-watch/\#TAs}, last checked March 5, 2024.} While the development is often seen as evidence of a change in the publishing market, it remains an open question whether this is a change for the better. 

The present paper investigates how publishers with TAs may adjust their pricing in response to a change in the number of publications in their journals, e.g., by a shift away from a publisher due to requirements by third-party funding, public boycotts, or generally changing publishing preferences of academics. I theoretically demonstrate that established publishers with many papers behind subscription paywalls may exploit the `must stock' dilemma in academic publishing: Universities must provide their researchers with at least the leading journals of every discipline. Even though intended to cause the opposite, the design of transformative agreements may actually increase prices in a market shifting towards open access. Furthermore, it may raise barriers to market entry, which will perpetuate the ruling position of the incumbent commercial publishers as long as these publishers possess a sufficient amount of paywalled publications. 

I particularly study competition between an established publisher with a transformative agreement and a fully/gold open-access publisher, as many stakeholders are dissatisfied with the large commercial incumbent publishing firms and hope for more competition, preferably from fully open-access providers. Furthermore, I apply my model under the assumption of fixed library budgets. Rising costs for providing scientific literature is a pressing issue for many mid-tier universities and even for leading institutions \parencite{Lariviere.2015}. 

The necessity to subscribe to the publishers' journal portfolios in the first place is rooted in the role of academic libraries as literature providers for faculty and students \parencite{Puehringer.2021, KlainGabbay.2019}. The high relevance of scholarly journals for the work and career of researchers turns the portfolios of large publishers into `must stock' products \parencite{Chone.2016}, which endows their owners with high market power. In one way or another, universities must provide them to their researchers by purchasing subscriptions.\footnote{One may get copies from print versions stored in libraries that are still available after a subscription cancellation, but it no longer applies to publications since a subscription has ended.} It enables the leading commercial publishers to extract enormous rents from the scientific community \parencite{Lariviere.2015}. Besides a fixed effort to maintain the server infrastructure, they face hardly any marginal costs of granting access to their repositories \parencite{Bergstrom.2014}, a common phenomenon in digital markets \parencite[see, e.g.,][]{Waldfogel.2017}.

An objection to the `must stock' proposition is the existence of predatory repositories that infringe copyrights to provide peer-reviewed research for free. Relying on such loopholes undermines the publishing system as a whole, according to \textcite{Himmelstein.2018}, or may even strengthen the subscription-based system \parencite{Strielkowski.2017} as in the entertainment industry, which open-access advocates do not intend. Last, official bodies like universities are not entitled to refer to predatory repositories but condemn these bypaths.

By analyzing the effects of the contract design of TAs on competition, this paper contributes to the economic analysis of science. The economic implications of open access have been discussed by, e.g., \textcite{Armstrong.2015, McCabe.2005}. \textcite{Bergstrom.2001, Jeon.2010, Haucap.2021} have pointed out the exaggerated economic position of journals -- and, by that, implicitly their publishers -- as they can be considered two-sided platforms \parencite[see, e.g.,][]{McCabe.2004}, which come along with the monopolization tendencies of network structures. While \textcite{Haucap.2021} and \textcite{Schmal.2023el} empirically study transformative agreements in Germany, they do not elaborate in detail on the economic implications for the publishers. \textcite{McCabe.2013} did the latter but compared subscription-based to open-access publishing in general, unrelated to transformative agreements. 

\textcite{Jahn.2024} finds that transformative agreements preserve the dominant position of the `big 3,' Elsevier, Springer Nature, and Wiley. The results correspond with \textcite{Butler.2023}, who identify the `big 3' as the largest benefactors of open access fees, whereas TAs are the most vital driver of open access income for Elsevier and Wiley. \textcite{McCabe.2014} demonstrate that open access contributes to more citations only for highly ranked journals -- this confirms why many researchers continue publishing in the journals of the legacy publishers. Even though \textcite{Ellison.2011} could show that scholars with very high reputation may skip the publishing process and may disseminate new work instead via unreviewed papers. \textcite{Schmal.2023} could show that gender differences exist, especially male researchers seek reputation. 

\textcite{Shu.2024}, among others, discuss the publishing fees for open access articles as a form of ``knowledge tax'' and find the publisher monopoly to prevail with large-scale open access adoption. \textcite{Rothfritz.2024} find that transformative agreements have become a permanent phenomenon instead of a transitory `bridge' to a fully open access world. Furthermore, they find that TAs with the leading three publishers are, on average, longer than those concluded with other publishers. 

The remainder of this paper is structured as follows. Section \ref{sec.model_pg} explains the theoretical modeling of this complex market setting. Section \ref{sec.analysis_pg} elaborates on how an incumbent publisher with a TA would optimally set the PAR fee and the resulting economic implications. Furthermore, I formalize the competition issue of the market entry of a fully open-access publisher. Section \ref{sec.conc_pg} concludes.

\section{Model}
\label{sec.model_pg}

The novelty of transformative agreements is the change in the revenue structure. Transformative agreements abolish the subscription fees and replace them with a -- on first sight -- payment structure fully independent of the publisher's repository of published papers. While TAs can take different forms, closest to the envisioned fully open access environment are contracts based on a single `publish-and-read' (PAR) fee: Affiliated institutions only pay a fixed fee per article published in an eligible journal with open access included by default. But the institutions also have full access to the partnering publisher's portfolio of paywalled publications \parencite{Hinchliffe.2019}. Analytically, the fee can be separated into a part that covers the costs of publishing the particular article (the `publish' component). A separate part covers the price of having access to the publisher's journal portfolio (the `read' component).\footnote{\label{fn.pg1} There exists strong suggestive evidence from the publishers that they actually decompose the PAR fee in the described way; see, e.g., the criticism from Elsevier that transformative agreements do not acknowledge the value of the stock of research provided by the publisher: \url{https://www.deutschlandfunk.de/konflikt-zwischen-hochschulen-und-elsevier-100.html}. Michael Böddeker interviewed Hannfired von Hindenburg (Elsevier representative), published on August 8, 2018.} Thus, I define the PAR fee $\phi$ as:
\begin{align}
\phi = \alpha \pi + (1-\alpha) \rho. \label{eq.phi1}
\end{align}
$\pi$ represents the `publish' part, $\rho$ the `read' part, and $\alpha \in [0,1]$ describes a weighting parameter of the two. Important for the relevance of $\rho$ is that the TA publisher continues to publish papers that require a subscription, e.g., papers of researchers from institutions without a TA. Only in this case can the publisher exploit a continuous flow of new publications for which reading access is preferable or even necessary.

In a PAR environment, the number of publications $N$ is the only source of revenue. Thus, the relative importance of the `publish' and the `read' part likely depends on it. Hence, I define the weighting parameter as $\alpha(N)$. For the publishing part, a relation to the number of publications is self-explanatory: The price will likely go up with a higher demand for journal space. In contrast, reading ($\rho$) only depends on the number of publications \emph{not} being part of a TA because all TA publications appear with open access. Thus, the portfolio of paywalled publications is exogenous. Therefore, I treat reading as a fixed term. This leads to a refinement of $\phi$ as follows:
\begin{align}
\phi = \phi(\pi, \rho, \alpha) = \alpha(N) \pi(N) + \big(1-\alpha(N)\bigr) \rho. \label{eq.phi2}
\end{align}

I assume the `publish part' function $\pi(N)$ of the PAR fee to be a twice differentiable, concave function as prices are arguably increasing with the relevance of a publisher captured by the number of publications, but the price cannot infinitely grow. Theoretically, the fee paid per paper could also decrease as there is fixed cost degression. However, \textcite{Budzinski.2020, Borrego.2023} show that additional exogenous confounders, such as reputation, affect the publication fees of journals. It is unlikely that publishers negotiating a TA abstain from this consideration. I assume the same behavior for $\alpha(N)$. For the sake of simplicity, I abstract from bargaining. Even though this is not entirely realistic, it helps to highlight the core issue of this paper: How transformative agreements may absorb library budgets regardless of the number of publications. By that, it can weaken existing fully open access publishers and raise barriers to entry, which harms competition. 

Particularly specific for academic publishing is that creating a demand function is problematic because buyers and users of the service (publication) differ. Plenty of stakeholders, such as university councils, university libraries, legislators, funding agencies, and science officials, participate directly or indirectly in price negotiations and product design, for example, what kinds of access they demand. In contrast, the core users of academic publications (both reading and publishing), namely the researchers, have little to no say in how the publication system is designed, how much is paid from whom to whom, and so on. While the researchers create a revenue through publishing their papers, universities or their libraries somewhat independently, negotiate with the publishers and buy the publication service as well as access to a publisher's publication portfolio. 

As an alternative, I examine what happens with the optimal setting of $\phi$ in response to a change in the number of publications $N$ and, as a consequence, how this affects the publisher's profit under a transformative agreement given that there continue to exist publications from institutions not covered by such a contract. I define the profit function of a publisher under a transformative agreement as follows:\footnote{For the sake of simplicity, I truncate the $i$ subscripts at the parameters $N$, $\phi$, $c$, and $F$ here. Furthermore, I abstract from revenue generated by submission fees because there is no plausible argument as to why they should change due to the switch to a transformative agreement.}
\begin{align}
\Pi_i = N \left(\phi - c \right) - F  \label{eq.profit1}
\end{align}
Revenue is generated by the number of published papers $N$ priced with the PAR fee $\phi$.\:$c$ captures the variable cost for each publication, $F$ represents fixed costs of maintaining the digital repository of papers given that in the digital age, variable costs of server usage are negligibly low. Second, I assume linear costs $c$ for every paper published. For the profit function, I only consider the revenues from the market in which the publisher concluded the transformative agreement, i.e., I abstract from subscription revenues generated elsewhere. Replacing $\phi$ in the publisher's profit function with the adjusted version provided in eq.~(\ref{eq.phi2}) leads to:
\begin{align}
\Pi_i &= N \bigl( \alpha(N)\pi(N) + (1-\alpha(N))\rho - c \bigr) - F \label{eq.profit2}
\end{align}

Optimally, I would build my analysis on a maximization problem of the publisher. Given that the PAR fee is fixed across journals and not paid based on the demand for published articles researchers want to read but on the number of publications, the publisher can only react to changes in the publication behavior of scientists at institutions with transformative agreements. In the German case of the `DEAL' contracts with Wiley and Springer Nature, even the initial setting of the PAR fee was fixed: The first TA contracts with Springer and Wiley were meant to be budget neutral, i.e., the PAR fee was based on the aggregated previous subscription costs of the participating institutions and divided by the number of expected publications. This resulted in a fee of 2,750 EUR per article \parencite{Kupferschmidt.2019}. Thus, the publishers could not freely bargain the PAR fee here. Subsequently, they would have to adjust it based on this starting value, which is exactly what I model in the following section.

\section{Pricing Considerations of the TA Publisher}
\label{sec.analysis_pg}
\vspace{-4mm}
%
In this section, I evaluate the optimal firm behavior in response to a change in publications. Just as explained, I assume $\phi$ to be already in equilibrium, i.e., in default of a clear demand function, I do not compute the optimal price of $\phi$ but the optimal reaction to an exogenous change in publications from the researchers.

\vspace{-2mm}
\subsection{Optimal setting of the PAR fee}
\label{ssec.fee}
The core elements of this model are, due to the design of transformative publish-and-read agreements, the exogenous number of publications $N$, and, of course, the PAR fee $\phi(\pi, \rho, \alpha)$, in which its components are functions of the publications by themselves. Hence, the TA publisher adjusts $\phi$ in reaction to a change in publications as follows:
\begin{align} \label{eq.phi_der1}
\frac{\partial \phi}{\partial N} = \frac{\partial \pi}{\partial N}\alpha(N) + \frac{\partial \alpha}{\partial N}\pi(N) - \frac{\partial \alpha}{\partial N}\rho
\end{align}	

The derivative in eq.~(\ref{eq.phi_der1}) consists of three parts. First, there is the adjustment of the analytic publishing part of the fee. This is positive given the increased demand for journal space in the publisher's outlets. Second, there is an internal adjustment of the weighting between the `publish' and the `read' part captured by the parameter $\alpha$. 
Modeling $\phi$ as $\phi = \alpha(N) \pi(N) + \big(1-\alpha(N)\bigr) \rho$ always leads to a corner solution of $\alpha \in \{0,1\}$. To study the effects of changes in the publication behavior of researchers on the profits of a TA, it needs a case distinction: Either the publication fee exceeds the library access fee ($\pi > \rho$) or vice versa.\footnote{I abstain from the fringe case of $\pi = \rho$ as any adjustment in $\alpha$ between both components of $\phi$ would be profit neutral.}

\begin{figure}[htbp]
\centering
\vspace{3mm}
\includegraphics[width=.7\linewidth]{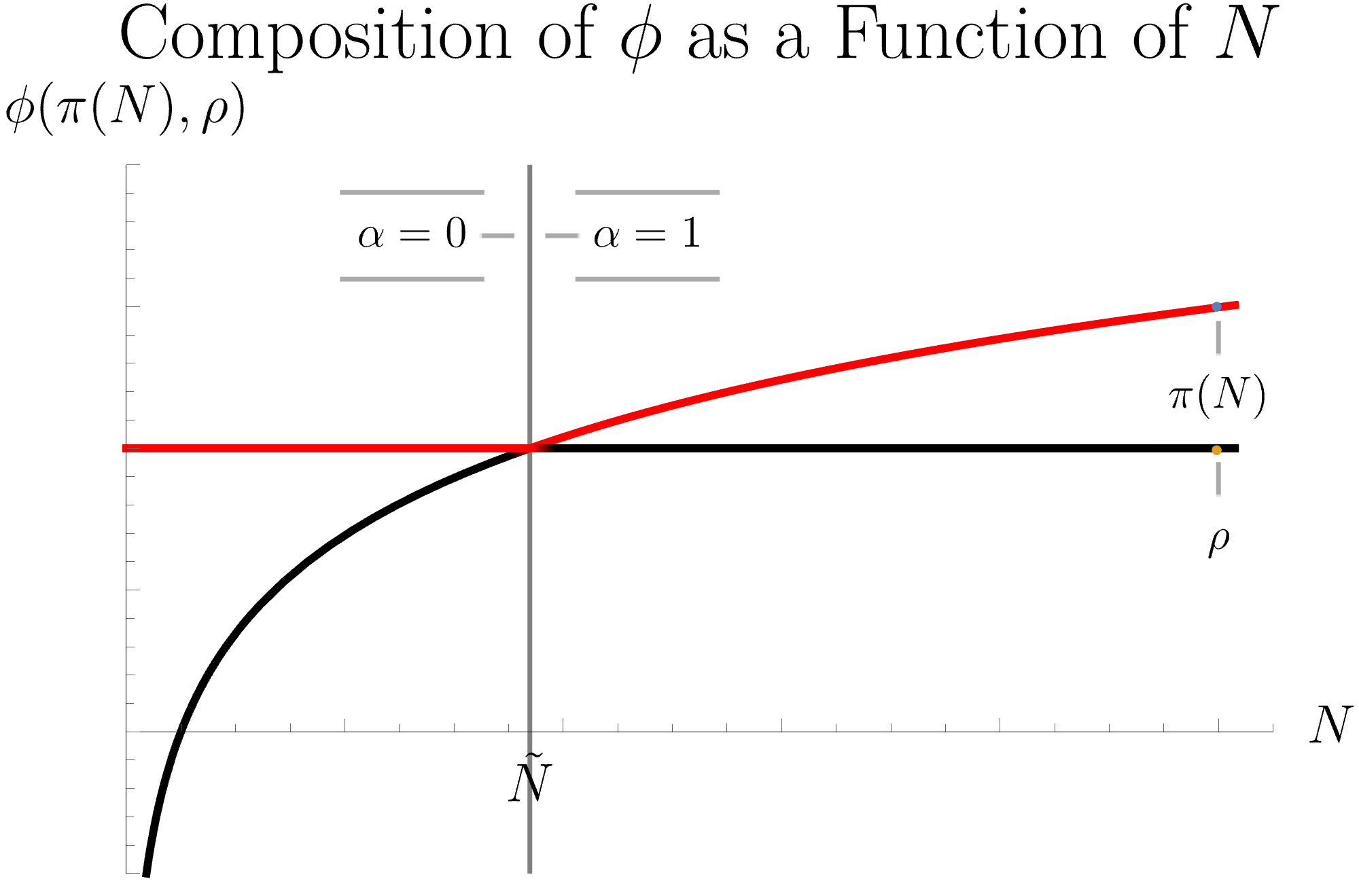}
\caption{Optimal price setting of the PAR fee as a function of $N$}
\label{fig.1}
\end{figure}

Figure \ref{fig.1} plots how $\phi$ is set as a function of $\pi(N)$ and $\rho$. The horizontal line captures $\rho$ as it is independent of the number of publications. $\pi(N)$ increases in $N$ as higher demand for journal space should increase its price. As long as $\pi(N)>\rho$ holds, the publication fee $\phi$ entirely consists of the publishing part (i.e., $\alpha = 1$). Once it switches to $\rho > \pi(N)$, the publisher focuses on its portfolio of past publications and uses this as pricing, i.e., $\phi = \rho$ and $\alpha = 0$. Hence, eq.~(\ref{eq.phi1}) essentially becomes $\phi = \max\{\pi(N),\rho\}$, which the red line captures in the plot.
The derivative of $\phi$ becomes $0$ for all $N<\tilde N$, where $\alpha$ switches as $\frac{\partial \alpha}{\partial N}=0$ for $\alpha = 1$. In contrast, $\frac{\partial \phi}{\partial N}$ becomes $\frac{\partial \pi}{\partial N}$ for $\pi(N) > \rho$, as $\alpha = 1$ holds, such that only the `publish' function $pi$ matters for $\phi$. Both is illustrated in Figure \ref{fig.1}. Thus, the publisher has a `safety net' for starkly decreasing interest in its journals (as a publication outlet) but can benefit from growing interest in them. This unilaterally beneficial fee setting for the TA publisher is summarized in Proposition \ref{prop.phi}.
\begin{proposition}\label{prop.phi}
With a transformative agreement, a publisher owning a portfolio of paywalled publications can monetize it by assigning the `read' part of the PAR fee a higher emphasis if its number of TA publications $N$ falls below the threshold $\tilde N$, i.e.,
\begin{align*}
\phi^* = 
\begin{cases}
\pi(N) & \mbox{if } N > \tilde N \rightarrow \alpha = 1\\
\rho & \mbox{if } N \le \tilde N \rightarrow \alpha = 0\\
\end{cases}
\end{align*} 
Even though the fee decreases in decreasing demand, this is stopped once $\tilde N$ is reached due to the switch to relying entirely on the `must stock' requirement embodied by the publisher's paywalled publications. \hspace*{0pt}\hfill $\blacksquare$
\end{proposition}

The pattern becomes more pronounced once one interprets $\rho$ as a function of $N$. Given the nature of supply and demand, assuming a concave function of the publishing part $\pi(N)$ is reasonable. However, no actual price can be inferred for the journal portfolio that is in the hands of such a publisher. First, the price is \emph{per se} no function of publications as the interest in the library is not related to publishing new papers. It might be indirectly the case via the reputation built in the past. However, reputation remains, at least in the short-run, the same regardless of the number of new publications. A price based on paper requests or downloads would be conceivable. But reading is not separately reimbursed anymore due to the design of the agreements based on the single \emph{publish-and-read} fee.     

\begin{figure}[htbp]
\vspace{3mm}
\centering
\includegraphics[width=.7\linewidth]{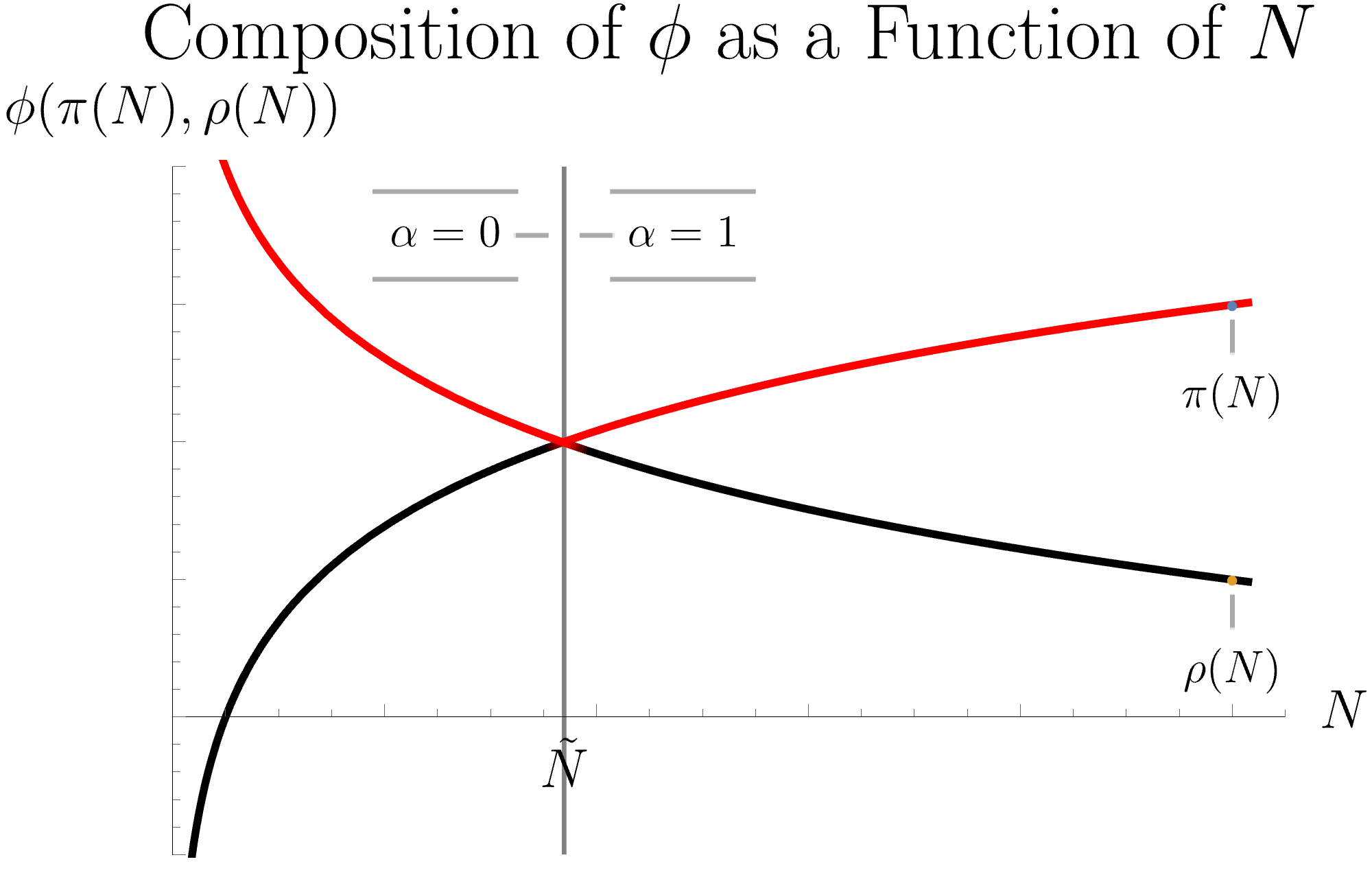}
\caption{Optimal price setting of the PAR fee as a function of $N$ with $\rho(N)$}
\label{fig.2}
\end{figure}

Nevertheless, the concave shape of the pricing function of the publication part of the PAR fee suggests a reverse function modeling for the `read' part from the publisher's perspective. In particular, a convex function with the properties $\frac{\partial \rho(N)}{\partial N}<0$ and $\frac{\partial^2 \rho(N)}{\partial N^2}\ge 0$ maximizes the publisher's marginal revenue as it holds for $\phi$ that $-\frac{\partial \phi}{\partial N}>0\:|\:\rho(N)>\pi(N)$. Figure \ref{fig.2} illustrates this.

With such a specification, the marginal revenue from a paper may actually \emph{increase} even if the number of publications of this publisher diminishes. The TA publisher possesses monopoly power over access to its journals, especially to the leading outlets to which the universities must gain access. Take the publisher Elsevier as an example. It hosts, among many others, the \emph{Journal of Development Economics}, the \emph{Journal of International Economics}, and the \emph{Journal of Monetary Economics}. As long as researchers from universities without such TAs continue to publish in these journals without open access, a stream of novel and, most likely, necessary research must be provided by the other universities to its researchers. Getting cut off from this body of research would be a significant challenge for any research institution.

A prominent example is Germany, as the country's universities and Elsevier parted ways over a dispute on how to design a joint transformative agreement. Consequently, researchers in Germany lost direct access to Elsevier in 2018 \parencite{Fraser.2021}. For several years, hardly any contract with Elsevier existed.
However, many universities offered their staff to order Elsevier publications for them on a case-by-case basis. Here, one can see that the German universities did not fully overcome the `must stock' challenge but, in turn, paid high amounts per article. According to the University of Cologne, it was approximately 35 EUR to access \emph{one} particular paper.\footnote{See  \url{https://web.archive.org/web/20221203193558/https://www.ub.uni-koeln.de/res/forms/article_elsevier/index_eng.html}. Copy from the Internet Archive, date of capture: December 3, 2022.} Thus, the `must stock \emph{per se}' problem has been turned into a `must stock \emph{per request}' for Elsevier. 

It underlines the central assumption of this paper and questions to which extent a hard exit from a leading incumbent publisher might be a viable alternative to the negotiation of a transformative agreement. Eventually, Elsevier and the German research institutions concluded a transformative agreement beginning in September 2023. From then on, access was restored, and every new paper appears with open access.\footnote{See \url{https://deal-konsortium.de/en/agreements/elsevier}, last checked September 18, 2024.} In the end, the forced cut-off could have strengthened the position of the libraries. However, the eventual agreement underlines that a final cut-off from such a publisher does not seem to be a realistic option, as the Germany-Elsevier dispute demonstrates.

\subsection{Profit of a monopolistic publisher}
\label{ssec.analysis_pg1}

The specific dynamics of the PAR fee directly affect a TA publisher's profit. In this subsection, I examine the reaction of a single monopolistic publisher with a TA to a change in the number of publications in its journals, i.e., how the already outlined adjustment of the PAR fee $\phi$ affects the publisher's profit. To do that, I compute the partial derivative of the publisher's profit function:
\begin{align}
\frac{\partial \Pi_i}{\partial N} &=\left(\phi - c \right) \label{eq.der1a} \\
\frac{\partial \Pi_i}{\partial N} &= \underbracket{\vphantom{\frac{\partial \pi}{\partial N}}\alpha \pi(N) + (1-\alpha(N))\rho - c}_{I}+  N \biggl(\: \underbracket{\alpha(N)\frac{\partial \pi}{\partial N}}_{II} + \underbracket{\frac{\partial \alpha}{\partial N}(\pi(N) - \rho)}_{III}\:\biggr) \label{eq.der1}
\end{align}
Equation (\ref{eq.der1a}) shows the `naive' derivative of the simplistic profit function shown in eq.~(\ref{eq.profit1}). Here, the publisher is a pure price taker.
Things get more involved once the TA publisher is not a price taker anymore but sensitive to changes in publications. The formerly exogenous PAR fee $\phi$ becomes $\phi(\pi, \rho, \alpha)$, a function of $\pi,\:\rho$, and $\alpha$, which again are depend on $N$. The respective derivative in eq.~(\ref{eq.der1}) consists of three parts: $I$ captures $\phi - c$ based on the definition of $\phi$ in eq.~(\ref{eq.phi2}). Hence, this is equivalent to the plain derivative in eq.~(\ref{eq.der1a}). $II$ and $III$ contain two additional fee shifters: $II$ is the change in the price of the `publish' part due to the higher demand, i.e., $\frac{\partial \pi}{\partial N}$. It is weakly positive because higher demand weakly increases $\phi$ for all publications. 
Part $III$ shows the adjustment of the weighting parameter $\alpha(N)$. It comes into play only when switching between 0 and 1.\footnote{Even though, in theory, the partial derivative is positive, and $\alpha$ could take any real value in the interval $[0,1]$. Furthermore, part $III$ contains the difference between the `publish' and the `read' part.} 

The TA publisher's marginal profit depends on the relationship between $\rho(N)$ and $\pi(N)$. The case distinction of Proposition \ref{prop_profit} depicts this. Due to the concave `publish' function and the linear or convex `read' function, a transformative agreement implies a comparatively comfortable situation for these established publishers. If the number of publications \emph{increases} (given $\pi(N) > \rho(N)$), the situation allows to adjust the overall price per publication, i.e., an increase that exceeds $\phi-c$. The latter would be the additional revenue without any price adjustment. On the other hand, a significant \emph{decrease} in the number of publications triggers the publisher's emphasis on its existing publication portfolio. Hence, the decrease in marginal revenue gets tempered by $\frac{\partial \rho}{\partial N}$, which becomes positive for a decrease in $N$. This leads to the following proposition:

\begin{proposition}\label{prop_profit}
The design of the PAR fee $\phi$ allows a TA publisher to increase its profit under increasing demand for publishing in its journals if $\alpha = 1$. Under decreasing demand, the decrease in its profit is dampened once it reaches the state of $\alpha = 0$: 
\begin{align*}
\frac{\partial \Pi}{\partial N} = 
\begin{cases}
\pi(N) - c + N\frac{\partial \pi}{\partial N} > \phi & \mbox{if } \alpha = 0 \rightarrow \pi(N) > \rho(N)\\
\rho(N) - c + N\frac{\partial \rho}{\partial N} < \phi & \mbox{if } \alpha = 1 \rightarrow  \pi(N) < \rho(N)
\end{cases} 
\end{align*}
Once the threshold $\tilde N$ is reached, the publisher is better off exploiting the `must stock' character of its paywalled publications. This can reduce the continued decrease in revenue caused by a decline in TA publications. \hspace*{0pt}\hfill $\blacksquare$
\end{proposition}

\begin{figure}[htbp]
\centering
\vspace{3mm}
\includegraphics[width=.7\linewidth]{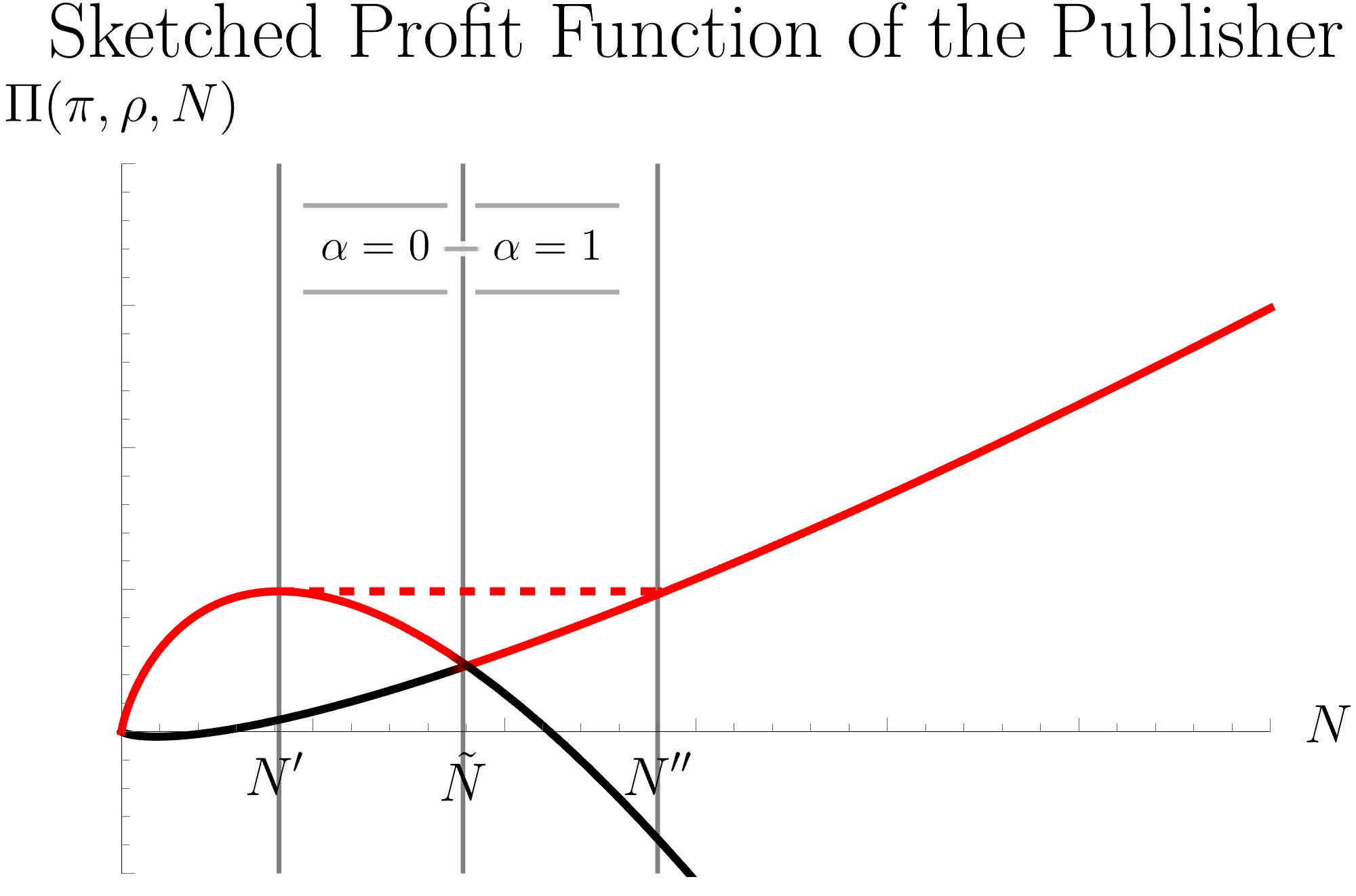}
\caption{Profits as a function of $N$ with $\rho(N)$}
\label{fig.3}
\end{figure}
Figure \ref{fig.3} shows how the profit function could look given the assumptions of a concave publish and a convex read part of the transformative publication fee $\phi$. Switching from $\alpha = 1$ to $\alpha = 0$ leads to a significant change in profits as the convexity of the read function $\rho(N)$ leads to an increase in the marginal and the overall profit. Of course, this cannot be sustained for very low values of $N$ as the decrease in the actual number of TA publications at some point outweighs the increase in $\rho$. 

Nevertheless, this situation may get worse from the perspective of the libraries. As the publisher wants to maximize profits, they likely set the profits based on the value of $\phi$ that maximizes its profit as described by the red dashed line in Figure \ref{fig.3}. In particular, it is reasonable for the publisher to entirely focus on its stock of papers much earlier and may sell this as the main asset when negotiating a TA. It stabilizes the publisher's profits in the interval $N \in [N', N'']$ with $N' \le \tilde N \le N''$ even though the number of publications is in decline. This emphasizes the prominent role of the existing repository of publications, as long as the TA publisher continues to publish papers requiring subscriptions from other countries or institutions without a transformative agreement.\footnote{One objection to the bargaining power of the incumbent publishers with a transformative agreement is that the relevance of the average publication diminishes over time. Hence, the older the stock of papers of a TA publisher, the less its power in negotiations. But the bargaining power remains strong as long as the inflow of restricted-access publications does not diminish.} 

\subsection{Competition with a full open access publisher}
\label{ssec.comp_pg}
In this section I add a fully open-access publisher as competitor. By definition, it possess neither an existing stock of publications nor can it build any. This scenario reflects reality: There are, on the one hand, large established firms such as Elsevier, Springer Nature, or Wiley. New market participants such as `Frontiers,' the `Public Library of Science' (PLoS), or the MDPI dedicated themselves solely to open access publishing.
For the fully open access publisher, I use the subscript $_{OA}$.\footnote{Of course, a TA publisher also publishes at least the TA-covered papers with open access.} 

Since the whole market consists of two publishers, the total number of publications is $\bar N = N_{TA} + N_{OA}$.
Secondly, I assume that university budgets are fixed.\footnote{This is not unrealistic because even well-endowed institutions such as Harvard University stated its incapability to pay for the expensive journal subscriptions, see, e.g., \textcite{Sample.2012}.} Hence, a library's budget $B$ is composed as $B = \phi_{TA} N_{TA} + \phi_{OA} N_{OA}$.\: 
$\phi_{TA}$ is the publication fee of the TA publisher (as defined in eq.~(\ref{eq.phi2})). $\phi_{OA}$ is the article publishing fee of the open access publisher. It consists only of a `publish' part because there exists no repository of restricted-access publications the OA publisher could sell. Hence, $\phi_{OA} = \pi_{OA}(N)$. 

Further, I assume for the `publish' functions $\pi_{TA}(N_{TA}) \ge  \pi_{OA}(N_{OA})$, because the incumbent TA publisher might have some first mover advantage, e.g., due to its grown reputation. Again, I assume a twice-differentiable concave function with $\frac{\partial \phi_{OA}(N)}{\partial N}>0$ and $\frac{\partial^2 \phi_{OA}(N)}{\partial N^2}\le 0$. Also, I assume the slope of the `publish' part function to be the same for both publishers, i.e., $\frac{\partial \pi_{TA}(N)}{\partial N} =\frac{\partial \pi_{OA}(N)}{\partial N}$. It is a simplification but a less severe constriction than assuming $\pi_{TA}(N) = \pi_{OA}(N)$. Furthermore, it is reasonable to assume that both publishers can \emph{change} their fees in the same way in response to a change in demand. While differences in market power usually affect prices \parencite{Budzinski.2020}, it is plausible that it only shifts the `publish part' upwards and does not change the functional form of the first derivative. 

Equipped with these prerequisites, I examine how the market changes if the \emph{relative} amount of publications shifts from one publisher to the other. This is because of the omnipresent calls to transform the academic publishing market. In the status quo, the TA publisher is likely to be in the $\pi(N) > \rho(N)$ case, i.e., it sets $\alpha = 1$ and focuses solely on the `publish' part. Any shift from the OA to the TA publisher is straightforward. An increase in $N_{TA}$ implies a growth in $\pi(N)$, by that, a growth in $\phi$ and a decrease in $\hat \phi$ as the demand for the OA publisher diminishes equivalently. 

The changes are more involved for the opposite -- and from many stakeholders desired -- case: What happens when researchers shift from the incumbent TA publishers to the new fully open-access market participants?\footnote{For the sake of tracktability, I assume a fixed number of publications. Notwithstanding the long-run growth in academic output \parencite{Bornmann.2021}, the assumption of a fixed amount eases the understanding of a shift in market shares, which captures relative variation. Constant growth in $N$ is easily conceivable, though, and would not change the results qualitatively.} Given $\bar N$, $\Delta N_{OA} = -\Delta N_{TA}$ must hold. Then, the partial derivative of $B$ by $N_{OA}$ becomes
\begin{align}
\frac{\partial B}{\partial N_{OA}} = \underbracket{\frac{\partial \phi_{OA}}{\partial N_{OA}}N_{OA}}_{I} - \underbracket{\frac{\partial \phi}{\partial N_{TA}}N_{TA}}_{II} + \underbracket{\vphantom{\frac{\partial B}{\partial N_{OA}}} \phi_{OA} - \phi_{TA} }_{III} \overset{!}{=} 0\:. \label{eq.noa1}
\end{align}
$I$ describes the change in the OA publisher's fee $\phi_{OA}$ multiplied by the number of publications. $II$ describes the equivalent for the TA publisher. Given a fixed amount of publications $\bar N$, a marginal shift \emph{towards} the OA publisher implies a shift \emph{away} from the TA publisher. $II$ describes the change in the fee $\phi_{TA}$ of the TA publisher weighted by the remaining TA publications $N_{TA}$. $III$ describes the marginal change in the budget by switching a marginal publication from the TA to the OA publisher. 

The derivative is set equal to zero as the budget is fixed. I want to demonstrate how the relative growth of the full OA publisher affects the fee it can charge under the restriction of the presence of a TA publisher. Accordingly, rearranging leads to:
\begin{align}
\underbracket{\frac{\partial \phi_{OA}}{\partial N_{OA}}N_{OA}}_{I} =  \underbracket{\frac{\partial \phi_{TA}}{\partial N_{TA}}N_{TA}}_{II} -  \underbracket{\vphantom{\frac{\partial B}{\partial N_{OA}}} \left( \phi_{OA} - \phi_{TA}\right)}_{III}\:,
\label{eq.noa2}
\end{align}
It is worth evaluating which sign the LHS of eq.~(\ref{eq.noa2}) takes under the condition of a fixed library budget $B$. Naturally, it should be positive, given the increasing demand for publications from the OA publisher. However, it depends on the change in the \emph{TA} publisher's pricing. It depends on the area of the $\phi_{TA}$ composition curve in which the TA publisher finds itself:

\underline{For $\alpha = 1$}, only the publish part matters, i.e., $\frac{\partial \phi_{TA}}{\partial N_{TA}}>0$. Given $\frac{\partial \pi_{TA}(N)}{\partial N} =\frac{\partial \pi_{OA}(N)}{\partial N}$, it must hold that $\phi_{TA} \ge \phi_{OA}$ as $\pi_{TA}(N) \ge \hat \pi_{OA}(N)$ since it is only supposed to vary in the intercept and, by assumption, the intercept of the TA publisher weakly exceeds the one of the fully OA publisher. This is an important restriction. But I consider it reasonable because the leading publishers are all hybrid TA publishers and arguably possess more market power than new market entrants. Thus, given a sufficiently high demand for TA publications ($\alpha = 1$), $II>0$ holds as for an increase in $N_{TA}$, the fee $\phi_{TA}$ of the TA publisher increases and vice versa. The sign of $III$ is not entirely clear but it has not much impact on the sign of the overall RHS as the fee differential for \emph{one} publication is unlikely to exceed the change multiplied by all publications.

\underline{For $\alpha = 0$}, the TA publisher fully focuses on its journal repository when setting the price for $\rho(N) = \phi_{TA}$. Hence, the fee does not react if $\rho$ is fixed but may even increase if it is a function of $N_{TA}$. One can already infer conceptually that a \emph{decrease} in publications triggering an \emph{increase} in the TA fee $\phi_{TA}$ puts pressure on the OA publisher's fee that cannot monetize any portfolio of paywalled publications. This is confirmed analytically when looking at the RHS of eq.~(\ref{eq.noa2}). Now, $II$ becomes:
\begin{align}
\frac{\partial \phi_{TA}}{\partial N_{TA}}N_{TA}
\begin{cases}
= 0 & \mbox{if } \alpha = 0 \:\land\: \frac{\partial \rho(N_{TA})}{\partial N_{TA}}= 0 \:\:(fixed\:\rho) \\
< 0 & \mbox{if } \alpha = 0 \: \land \: \frac{\partial \rho(N_{TA})}{\partial N_{TA}}<0 \:\:(convex\:\rho\:function)
\end{cases}
\end{align}
Crucially, the case distinction for the TA publisher also affects the fully open-access publisher, which leads to Proposition \ref{prop_taoa}:

\begin{proposition} \label{prop_taoa}
The change in the fee chargeable by the fully open-access publisher depends not only on the demand for its own journals but also on the demand for publishing in the outlets of a publisher with a transformative agreement.
\begin{align*}
\frac{\partial \phi_{OA}}{\partial N_{OA}}N_{OA}
\begin{cases}
> 0 & \mbox{if } \alpha = 1 \\
\approx 0 & \mbox{if } \alpha = 0 \:\land\: \frac{\partial \rho(N_{TA})}{\partial N_{TA}}= 0\\
< 0 & \mbox{if } \alpha = 0 \:\land\: \frac{\partial \rho(N_{TA})}{\partial N_{TA}}<0 
\end{cases}
\end{align*}
A fully open-access publisher does not necessarily benefit from publication growth in its own journals when it operates in a market environment with an incumbent TA publisher that continues to publish restricted-access publications. \hspace*{0pt}\hfill $\blacksquare$
\end{proposition}
Due to part $III$, $\frac{\partial \phi_{OA}}{\partial N_{OA}}N_{OA}\approx 0$ holds for a fixed $\rho$ as the small differential in publication fees between OA and TA publisher for the marginal publication is not necessarily equal to zero but may differ slightly. Nevertheless, the main takeaway of Proposition \ref{prop_taoa} is that, under the condition of $\alpha = 0$, i.e., a rather `weak' TA publisher in terms of articles published, the change in the fee $\phi_{OA}$ of the full open-access publisher becomes weakly negative even if the demand for fully open-access publications actually \emph{increases}. It is due to the `must stock' problem the university libraries encounter with transformative publishers that still own a large stock of publications from the past and continue to publish papers behind paywalls. Thus, with fixed budgets, the library needs to push down the publication fee of the fully open-access publisher.\footnote{This scenario currently holds, but only as long as not all countries/institutions around the globe have closed transformative agreements, which would turn the TA into an OA publisher.}

\section{Outlook}
\label{sec.conc_pg}

Academic institutions and publishers around the globe continue to negotiate transformative agreements as an industrial policy to turn the portfolios of the large legacy publishers into open access outlets. The main goal of the universities is to substantially increase open-access publications to reduce the necessity of (expensive) journal subscriptions. At the same time, the payment streams shall change from paying for subscriptions to paying for the free availability of a paper. In addition, it is at least an implicit aim to reduce the money paid by universities or their libraries to the publishers. 

The present analysis has shown that transformative agreements centered around one single `publish-and-read' fee that decouples the access to a TA publisher's existing portfolio and the payment for it may perpetuate publishers' profits for a long time if not extend them. The results demonstrate that the mechanics of such transformative agreements raise entry barriers for new competing publishers or lower the budget for existing gold open access publishers, given the limited availability of additional funds to university libraries. Using contracts to lock out competitors is not new \parencite{Aghion.1987}, but up to this aspect has been missing in the discourse on academic publishing. Put differently, libraries endanger the early adopters by incorporating the legacy publishers into the open access transformation -- those who were reluctant to change in light of the high subscription revenues they want to protect. 

Overall, this paper's results raise questions about howhich transformative agreements are `transformative.' Other than intended, they may strengthen the position of the incumbent publishers instead of opening the market to new entrants and further straining library budgets. The crucial point in overcoming the status quo is fostering competition rather than open access at any cost. Without changes in the way contracts are designed, it could happen that Mallet du Pan's judgment of the French Revolution may also become applicable to the open access transformation of academic publishing: La révolution dévore ses enfants.

\vspace{10mm}
\begin{singlespace}
\printbibliography

@article{MuellerLanger.2018,
author = {Mueller-Langer, Frank and Watt, Richard},
title = {How Many More Cites is a \$3,000 Open Access Fee Buying Yoy? Empirical Evidence from a Natural Experiment},
journal = {Economic Inquiry},
volume = {56},
number = {2},
pages = {931-954},
year = {2018}
}

@article{Rothfritz.2024,
	author = {Rothfritz, Laura and Herb, Ulrich and Schmal, W Benedikt},
	title = {Trapped in Transformative Agreements? A Multifaceted Analysis of $>$1,000 Contracts},
	journal = {},
	volume = {},
	issue = {},
	year = {2024},
	note = {mimeo}
}

@article{Shu.2024,
  title={The oligopoly of open access publishing},
  author={Shu, Fei and Larivi{\`e}re, Vincent},
  journal={Scientometrics},
  volume={129},
  pages={519--536},
  year={2024},
  publisher={Springer}
}

@article{Butler.2023,
    author = {Butler, Leigh-Ann and Matthias, Lisa and Simard, Marc-André and Mongeon, Philippe and Haustein, Stefanie},
    title = "{The oligopoly’s shift to open access: How the big five academic publishers profit from article processing charges}",
    journal = {Quantitative Science Studies},
    volume = {4},
    number = {4},
    pages = {778-799},
    year = {2023},
    month = {11}
}

@article{Jahn.2024,
  title={How open are hybrid journals included in transformative agreements?},
  author={Jahn, Najko},
  journal={arXiv preprint arXiv:2402.18255},
  year={2024}
}

@article{Laakso.2016,
 author = {Laakso, Mikael and Bj{\"o}rk, Bo-Christer},
 year = {2016},
 title = {{Hybrid open access---A longitudinal study}},
 %url = {\url{https://www.sciencedirect.com/science/article/pii/S1751157716301523}},
 pages = {919--932},
 pagination = {page},
 volume = {10},
 number = {4},
 issn = {1751-1577},
 journal = {{Journal of Informetrics}},
 doi = {\url{10.1016/j.joi.2016.08.002}}
}

@article{KlainGabbay.2019,
 author = {{Klain Gabbay}, Liat and Shoham, Snunith},
 year = {2019},
 title = {{The role of academic libraries in research and teaching}},
 pages = {721--736},
 pagination = {page},
 volume = {51},
 number = {3},
 issn = {0961-0006},
 journal = {{Journal of Librarianship and Information Science}},
 doi = {\url{10.1177/0961000617742462}}
}

@article{Bergstrom.2014,
 author = {Bergstrom, Theodore C. and Courant, Paul N. and McAfee, R. Preston and Williams, Michael A.},
 year = {2014},
 title = {{Evaluating big deal journal bundles}},
 pages = {9425--9430},
 pagination = {page},
 volume = {111},
 number = {26},
 journal = {{Proceedings of the National Academy of Sciences}},
 doi = {\url{10.1073/pnas.1403006111}}
}

@article{Kupferschmidt.2019,
 author = {Kupferschmidt, Kai},
 year = {2019},
 title = {{Deal reveals what scientists in Germany are paying for open access}},
 issn = {1095-9203},
 journal = {{Science}},
 url = {https://doi.org/10.1126/science.aax1064},
 note = {Published February 21, 2019, last checked June 28, 2023.},
 doi = {10.1126/science.aax1064}
}

@article{Sample.2012,
 abstract = {University wants scientists to make their research open access and resign from publications that keep articles behind paywalls{\textless}/p{\textgreater}},
 author = {Sample, Ian},
 year = {2012},
 %year = {2012-04-24},
 title = {{Harvard University says it can't afford journal publishers' prices}},
 url = {https://www.theguardian.com/science/2012/apr/24/harvard-university-journal-publishers-prices},
 note={Published April 24, 2012, last checked June 26, 2023},
 %urldate = {June 26, 2023},
 journal = {{The Guardian}}
}

@article{Bornmann.2021,
 author = {Bornmann, Lutz and Haunschild, Robin and Mutz, R{\"u}diger},
 year = {2021},
 title = {{Growth rates of modern science: a latent piecewise growth curve approach to model publication numbers from established and new literature databases}},
 %url = {\url{https://www.nature.com/articles/s41599-021-00903-w}},
 pages = {1--15},
 pagination = {page},
 volume = {8},
 number = {1},
 issn = {2662-9992},
 journal = {{Humanities and Social Sciences Communications}},
 doi = {\url{10.1057/s41599-021-00903-w}}
}

@article{Schmal.2023a,
 author = {Schmal, W. Benedikt},
 year = {2023},
 title = {{The X Factor: Open Access, New Journals, and Incumbent Competitors}},
 journal = {MSI Discussion Paper No.~2307},
 note = {KU Leuven}
}

@article{Schmal.2023el,
 author = {Schmal, W. Benedikt},
 year = {2024},
 title = {{How Transformative are Transformative Agreements? Evidence from Germany Across Disciplines}},
 journal = {Scientometrics},
 volume = {129},
 pages = {1863–-1889}
}

@article{Schmal.2023,
 author = {Schmal, W Benedikt and Haucap, Justus and Knoke, Leon},
 year = {2023},
 title = {{The role of gender and coauthors in academic publication behavior}},
 journal = {Research Policy},
 volume = {52},
 issue = {10}
}

@article{Haucap.2021,
 author = {Haucap, Justus and Moshgbar, Nima and Schmal, W. Benedikt},
 year = {2021},
 title = {{The impact of the German 'DEAL' on competition in the academic publishing market}},
 pages = {2027--2049},
 pagination = {page},
 volume = {42},
 number = {8},
 issn = {0143-6570},
 journal = {{Managerial and Decision Economics}},
 doi = {\url{10.1002/mde.3493}}
}

@article{Budzinski.2020,
 author = {Budzinski, Oliver and Grebel, Thomas and Wolling, Jens and Zhang, Xijie},
 year = {2020},
 title = {{Drivers of article processing charges in open access}},
 %url = {\url{https://link.springer.com/article/10.1007/s11192-020-03578-3}},
 pages = {2185--2206},
 pagination = {page},
 volume = {124},
 number = {3},
 issn = {1588-2861},
 journal = {{Scientometrics}},
 doi = {\url{10.1007/s11192-020-03578-3}}
}

@article{McCabe.2013,
  title={Open access versus traditional journal pricing: Using a simple “platform market” model to understand which will win (and which should)},
  author={McCabe, Mark J. and Snyder, Christopher M. and Fagin, Anna},
  journal={The Journal of Academic Librarianship},
  volume={39},
  number={1},
  pages={11--19},
  year={2013},
  publisher={Elsevier}
}

@techreport{McCabe.2004,
  title={The economics of open-access journals},
  author={McCabe, Mark J. and Snyder, Christopher M.},
  year={2004},
  note = {Working Paper 193},
  institution={George J.~Stigler Center for the Study of the Economy and the State at the Unversity of Chicago}
}

@article{Fraser.2021,
  title={No deal: German researchers’ publishing and citing behaviors after Big Deal negotiations with Elsevier},
  author={Fraser, Nicholas and Hobert, Anne and Jahn, Najko and Mayr, Philipp and Peters, Isabella},
  journal={Quantitative Science Studies},
  volume={4},
  number={2},
  pages={325--352},
  year={2023},
  publisher={MIT Press}
}

@article{Borrego.2023,
author = {Borrego, Ángel},
title = {Article processing charges for open access journal publishing: A review},
journal = {Learned Publishing},
volume = {36},
number = {3},
pages = {359-378},
year = {2023}
}

@misc{Hinchliffe.2019,
 abstract = {Read-and-publish? Publish-and-read? A primer on transformative agreements},
 author = {Hinchliffe, Lisa Janicke},
 year = {2019},
 title = {{Read-and-publish? Publish-and-read? A primer on transformative agreements}},
 url = {https://scholarlykitchen.sspnet.org/2019/04/23/transformative-agreements/},
 %urldate = {2023-02-27},
 note = {Published April 23, 2019, last checked June 26, 2023},
 publisher = {Scholarly Kitchen}
}

@article{Waldfogel.2017,
  title={How digitization has created a golden age of music, movies, books, and television},
  author={Waldfogel, Joel},
  journal={Journal of Economic Perspectives},
  volume={31},
  number={3},
  pages={195--214},
  year={2017},
  publisher={American Economic Association 2014 Broadway, Suite 305, Nashville, TN 37203-2418}
}

@article{Bergstrom.2001,
  title={Free labor for costly journals?},
  author={Bergstrom, Theodore C},
  journal={Journal of Economic Perspectives},
  volume={15},
  number={4},
  pages={183--198},
  year={2001},
  publisher={American Economic Association}
 }

@article{Armstrong.2015,
 author = {Armstrong, Mark},
 year = {2015},
 title = {{Opening Access to Research}},
 pages = {F1-F30},
 pagination = {page},
 volume = {125},
 number = {586},
 issn = {0013-0133},
 journal = {{The Economic Journal}},
 doi = {\url{10.1111/ecoj.12254}}
}

@article{Jeon.2010,
 author = {Jeon, Doh-Shin and Rochet, Jean-Charles},
 year = {2010},
 title = {{The Pricing of Academic Journals: A Two-Sided Market Perspective}},
 pages = {222--255},
 pagination = {page},
 volume = {2},
 number = {2},
 issn = {1945-7669},
 journal = {{American Economic Journal: Microeconomics}},
 doi = {\url{10.1257/mic.2.2.222}}
}

@article{Aghion.1987,
 author = {Aghion, Philippe and Bolton, Patrick},
 year = {1987},
 title = {{Contracts as a Barrier to Entry}},
 %url = {\url{http://www.jstor.org/stable/1804102}},
 pages = {388--401},
 pagination = {page},
 volume = {77},
 number = {3},
 issn = {00028282},
 journal = {{The American Economic Review}}
}

@article{Puehringer.2021,
 author = {Puehringer, Stephan and Rath, Johanna and Griesebner, Teresa},
 year = {2021},
 title = {{The political economy of academic publishing: On the commodification of a public good}},
 %url = {\url{https://journals.plos.org/plosone/article?id=10.1371/journal.pone.0253226}},
 pages = {e0253226},
 pagination = {page},
 volume = {16},
 number = {6},
 issn = {1932-6203},
 journal = {{PLoS one}},
 doi = {\url{10.1371/journal.pone.0253226}}
}

@article{Chone.2016,
 author = {Chon{\'e}, Philippe and Linnemer, Laurent},
 year = {2016},
 title = {{Nonlinear pricing and exclusion:II. Must-stock products}},
 pages = {631--660},
 pagination = {page},
 volume = {47},
 number = {3},
 issn = {1756-2171},
 journal = {{The RAND Journal of Economics}},
 doi = {\url{10.1111/1756-2171.12138}}
}

@article{Himmelstein.2018,
 author = {Himmelstein, Daniel S. and Romero, Ariel Rodriguez and Levernier, Jacob G. and Munro, Thomas Anthony and Mclaughlin, Stephen Reid and Tzovaras, Bastian Greshake and Greene, Casey S.},
 year = {2018},
 title = {{Research: Sci-Hub provides access to nearly all scholarly literature}},
 journal = {{eLife}},
 colume = {7},
 note = {e32822}
}

@article{Strielkowski.2017,
 author = {Strielkowski, Wadim},
 year = {2017},
 title = {{Will the rise of Sci-Hub pave the road for the subscription-based access to publishing databases?}},
 pages = {540--542},
 volume = {33},
 number = {5},
 issn = {0266-6669},
 journal = {{Information Development}},
}

@article{Lariviere.2015,
  title={The oligopoly of academic publishers in the digital era},
  author={Larivi{\`e}re, Vincent and Haustein, Stefanie and Mongeon, Philippe},
  journal={PloS one},
  volume={10},
  number={6},
  pages={e0127502},
  year={2015},
  publisher={Public Library of Science}
}

@article{Dewatripont.2007,
 author = {Dewatripont, Mathias and Ginsburgh, Victor and Legros, Patrick and Walckiers, Alexis},
 year = {2007},
 title = {Pricing of Scientific Journals and Market Power},
 pages = {400--410},
 volume = {5},
 number = {2-3},
 issn = {1542-4766},
 journal = {Journal of the European Economic Association},
 doi = {10.1162/jeea.2007.5.2-3.400}
}

@article{McCabe.2005,
 author = {McCabe, Mark J. and Snyder, Christopher M.},
 year = {2005},
 title = {Open Access and Academic Journal Quality},
 pages = {453--458},
 volume = {95},
 number = {2},
 issn = {0002-8282},
 journal = {American Economic Review}
}

@article{McCabe.2014,
 author = {McCabe, Mark J. and Snyder, Christopher M.},
 year = {2014},
 title = {Identifying the effect of open access on citations using a panel of science journals},
 pages = {1284--1300},
 volume = {52},
 number = {4},
 issn = {0095-2583},
 journal = {Economic Inquiry}
}

@article{Ellison.2011,
 author = {Ellison, Glenn},
 year = {2011},
 title = {Is peer review in decline?},
 pages = {635--657},
 volume = {49},
 number = {3},
 issn = {0095-2583},
 journal = {Economic Inquiry},
 doi = {10.1111/j.1465-7295.2010.00261.x}
}
\end{singlespace}
\pagebreak

\end{doublespace}
\end{document}